\documentclass{ptapap}

\author{Anna Jacyszyn-Dobrzeniecka}[OAUW]
\author{the OGLE Team}
\affil[OAUW]{
Warsaw University Observatory \\
Al. Ujazdowskie 4, 00-478 Warszawa, Poland
}
  
\title{Three-Dimensional Structure of the Magellanic System}

\begin{document}

\maketitle

\begin{abstract}

We have determined the three-dimensional structure of the Magellanic Clouds and Magellanic Bridge using over $9\,000$ Classical Cepheids (CCs) and almost $23\,000$ RR~Lyrae (RRL) stars from the fourth phase of the OGLE project.

For the CCs we calculated distances based on period-luminosity relations. CCs in the LMC are situated mainly in the bar that shows no offset from the plane of the LMC. The northern arm is also very prominent with an additional smaller arm. Both are located closer to us than the entire sample. The SMC has a non-planar structure that can be described as an ellipsoid extended almost along the line of sight. We also classified nine of our CCs as Magellanic Bridge objects. These Cepheids show a large spread in three-dimensions.

For the RRL stars, we calculated distances based on photometric metallicities and theoretical relations. Both Magellanic Clouds revealed a very regular structure. We fitted triaxial ellipsoids to our LMC and SMC samples. In the LMC we noticed a very prominent, non-physical blend-artifact that prevented us from analyzing the central parts of this galaxy. We do not see any evidence of a bridge-like connection between the Magellanic Clouds.

\end{abstract}

\section{Introduction}

The goal of our study was to analyze the three-dimensional structure of the Magellanic Clouds using the OGLE Collection of Variable Stars \citep{Soszynski2015,Soszynski2016}. This collection is based on the fourth phase of the OGLE project (OGLE-IV) \citep{Udalski2015} that covers a vast area in the Magellanic System -- over 650 square degrees. This was the first time we were able to see a full picture of the Clouds using classical pulsators. Previous studies were based on OGLE-III that was monitoring an incomparably smaller area.

\section{Classical Cepheids in the Magellanic System}

\begin{figure}[htb]
	\includegraphics[width=\textwidth]{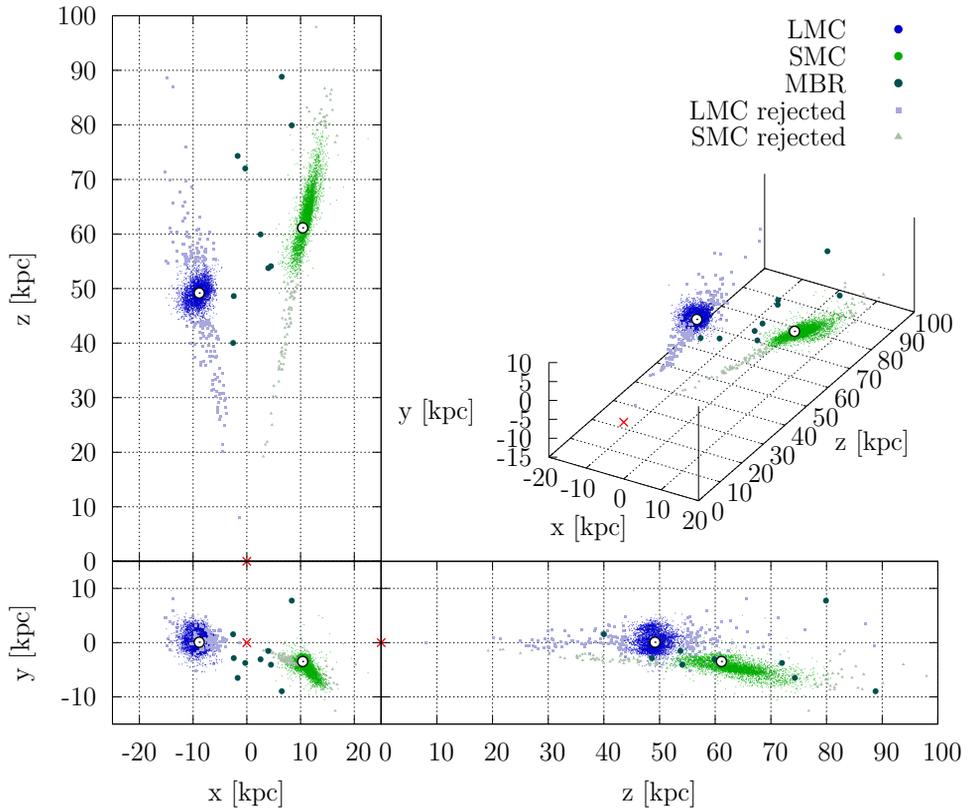}
	\caption{Three-dimensional map from \citet{AJD2016} showing classical Cepheids in the Magellanic System in Cartesian coordinates. Blue dots represent the LMC, green dots the SMC, and large dark teal dots the MBR. Gray points show the $3\sigma$ outliers rejected in the $P-L$ fitting procedure.}
	\label{fig:cc}
\end{figure}

Our final sample consisted of over $9\,000$ classical Cepheids (CCs). To calculate individual distances to each Cepheid, we fitted a period--luminosity ($P-L$) relation. We assumed that the fitted line corresponds to the mean LMC distance \citep{Pietrzynski2013}. Fig.~\ref{fig:cc} shows the final three-dimensional CC distribution in the Magellanic System.

\subsection{CCs in the LMC}

The CC distribution in the LMC in three dimensions clearly shows that these stars are clumped. We can see a very prominent bar and northern arm. Also, this distribution shows the inclination of the LMC disk. Most CCs are located in the central parts of the LMC.

We have divided stars into subgroups corresponding to different structures to make a statistical analysis. These structures were the bar, northern arm, and two additional structures in the south. The bar and northern arm were additionally subdivided into two parts.

The bar was divided into the western and eastern parts. The former is usually referred to as `the bar' in the literature (also `classical bar') and is the brightest part of the entire bar, while the latter is connecting the bar and the northern arm. We have carefully analyzed both parts of the bar and found that they are consistent in terms of CC locations and ages. This `new' bar shows no offset from the LMC disk. For age estimates in the LMC and SMC, we used period-age relations from \citet{Bono2005}.

The northern arm was divided into the main part (primary), and an additional smaller arm located towards the northern end of the primary arm. The entire northern arm shows an offset of about 0.5\,kpc and is located closer to us than the mean LMC distance.

\subsection{CCs in the SMC}

The CC distribution in the SMC is more regular than in the LMC. This distribution is very elongated, almost along the line of sight, and cannot be described as a disk. It is rather ellipsoidal.

We have found two off-axis substructures in the SMC. These structures are not as prominent as those in the LMC. The northern SMC substructure is located closer to us and it is younger, while the southwestern SMC substructure is located farther and is older on average.

\subsection{CCs in the MBR}

Based on the three-dimensional locations, we have classified 9 CCs as Magellanic Bridge (MBR) members. Their on-sky locations are well correlated with a young star distribution from \citet{Skowron2014} and a neutral hydrogen distribution \citep[]{Kalberla2005}. All the CCs are very young with ages less than 300\,Myr. Five of the best correlated ones seem to form an on-sky connection between the Clouds.

Their three-dimensional locations are not as well correlated. Only four of nine CCs may form a physical connection between the Clouds. The closest MBR CC is closer than any LMC CC while the farthest is farther than any SMC CC.

\section{RR~Lyrae Stars in the Magellanic System}

\begin{figure}[htb]
\includegraphics[width=\textwidth]{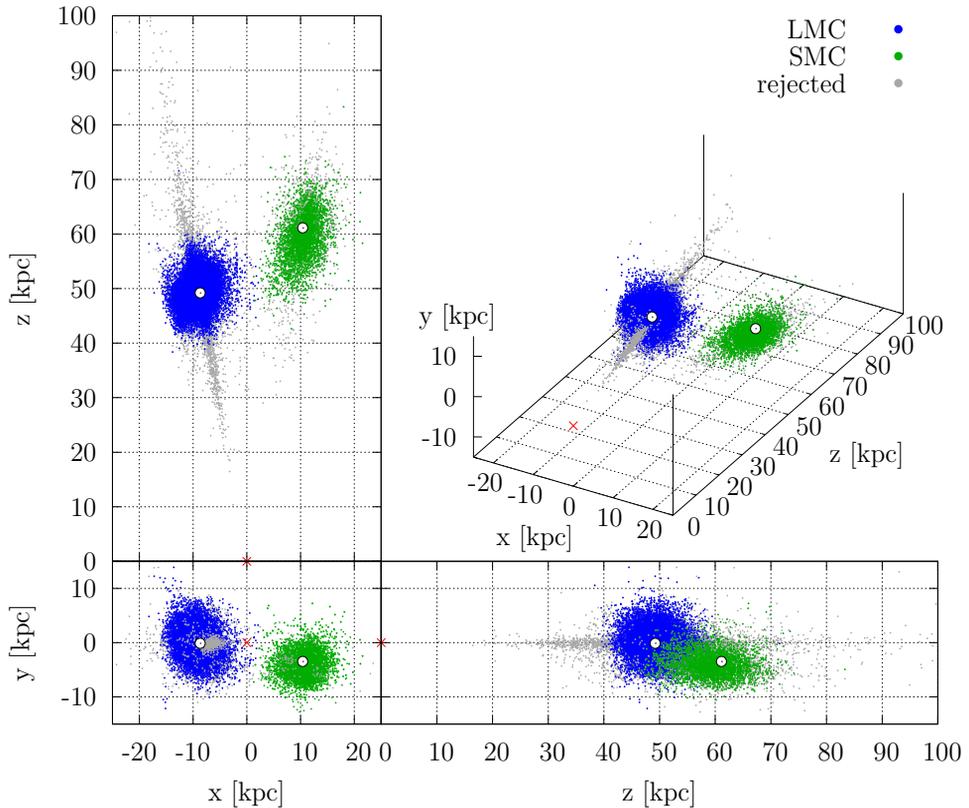}
\caption{The RRL stars from \citet{AJD2017} in the Magellanic System in Cartesian coordinates. The LMC stars are marked with blue dots, while the SMC stars -- with green dots. Additionally, all the rejected RRL stars are shown in gray.}
\label{fig:rrl}
\end{figure}

The second part of our project was the analysis of a three-dimensional distribution of RR~Lyrae (RRL) stars in the Magellanic System. Here we used a different method to calculate individual distances. We used photometric metallicity \citep[see][]{Skowron2016} and combined it with a period to obtain absolute magnitudes. Then, having visual and absolute magnitudes (reddening-independent Wesenheit index), we could easily calculate distances. Our total sample consisted of almost $23\,000$ RRL stars. Fig.~\ref{fig:rrl} shows the three-dimensional distribution of these objects in the Magellanic System.

\subsection{RRL Stars in the LMC}

RRL stars in the LMC in three dimensions show a very regular shape, and we do not see any evidence of substructures or irregularities. There is only one exception to this: a non-physical blend artifact that seems to be drawn out of the LMC center towards and apart from the observer along the line of sight. This structure is caused by the very intense blending and crowding effects in the LMC center, and it prevented us from studying this area in detail. We tried to remove it by making cuts in different parameter spaces, but we were unable.

To the LMC RRL stars distribution, we fitted triaxial ellipsoids. The innermost ellipsoids are probably affected by the blend artifact and thus their parameters may not be physical. The outer ellipsoids tend to be more twisted towards the SMC.

\subsection{RRL Stars in the SMC}

RRL stars in the SMC have an even more regular shape than in the LMC with no additional substructures or irregularities. We also fitted triaxial ellipsoids. Virtually all of them have the same shape (axes ratio). Again, the outermost ellipsoids are more twisted towards the LMC.

\subsection{RRL Stars in the MBR}

In the area of the Magellanic Bridge we can clearly see the two halos of the Magellanic Clouds overlapping. We do not see any evidence of an additional connection between the Clouds. The number of stars is too low to state that there is such a connection, and a statistical analysis is needed. We were unable to perform such analysis with our data because of the LMC halo reaching farther than OGLE-IV sky coverage. We are currently working on modelling the LMC halo using ellipsoids that we have fitted to the data (Jacyszyn-Dobrzeniecka et al. in prep.). There is no such problem with the SMC halo as we can see it entirely.

Recently, \citet{Belokurov2017} analyzed OGLE-IV RRL stars and found an actual bridge connecting the two Magellanic Clouds. We tried to reproduce their result -- we used exactly the same method as they did to calculate metallicities and distance moduli. Even though we set a similar scale, range, bin sizes, and colour on our plot, we do not see the connection. We also plotted the data with different parameters (different metallicity and distance modulus cuts and different bin size and range), and none of our plots shows the connection. This has confirmed results from our paper that there actually is no evidence of a connection between the Magellanic Clouds.

\acknowledgements{A.M.J.-D. is supported by the Polish Ministry of Science and Higher Education under ``Diamond Grant'' No. 0148/DIA/2014/43.}

\bibliographystyle{ptapap}
\bibliography{AJD}

\begin{thebibliography}{11}
\providecommand{\natexlab}[1]{#1}
\providecommand{\url}[1]{\texttt{#1}}
\providecommand{\urlprefix}{URL }
\providecommand{\eprint}[2][]{\url{#2}}

\bibitem[{{Belokurov} et~al.(2017)}]{Belokurov2017}
{Belokurov}, V., et~al., \emph{{Clouds, Streams and Bridges. Redrawing the
  blueprint of the Magellanic System with Gaia DR1}}, \emph{\mnras}
  \textbf{466}, 4711 (2017), \eprint{1611.04614}

\bibitem[{{Bono} et~al.(2005)}]{Bono2005}
{Bono}, G., et~al., \emph{{Classical Cepheid Pulsation Models. X. The
  Period-Age Relation}}, \emph{\apj} \textbf{621}, 966 (2005),
  \eprint{astro-ph/0411756}

\bibitem[{{Jacyszyn-Dobrzeniecka} et~al.(2016)}]{AJD2016}
{Jacyszyn-Dobrzeniecka}, A.~M., et~al., \emph{{OGLE-ing the Magellanic System:
  Three-Dimensional Structure of the Clouds and the Bridge Using Classical
  Cepheids}}, \emph{\actaa} \textbf{66}, 149 (2016), \eprint{1602.09141}

\bibitem[{{Jacyszyn-Dobrzeniecka} et~al.(2017)}]{AJD2017}
{Jacyszyn-Dobrzeniecka}, A.~M., et~al., \emph{{OGLE-ing the Magellanic System:
  Three-Dimensional Structure of the Clouds and the Bridge using RR Lyrae
  Stars}}, \emph{\actaa} \textbf{67}, 1 (2017), \eprint{1611.02709}

\bibitem[{{Kalberla} et~al.(2005)}]{Kalberla2005}
{Kalberla}, P.~M.~W., et~al., \emph{{The Leiden/Argentine/Bonn (LAB) Survey of
  Galactic HI. Final data release of the combined LDS and IAR surveys with
  improved stray-radiation corrections}}, \emph{\aap} \textbf{440}, 775 (2005),
  \eprint{astro-ph/0504140}

\bibitem[{{Pietrzy{\'n}ski} et~al.(2013)}]{Pietrzynski2013}
{Pietrzy{\'n}ski}, G., et~al., \emph{{An eclipsing-binary distance to the Large
  Magellanic Cloud accurate to two per cent}}, \emph{\nat} \textbf{495}, 76
  (2013), \eprint{1303.2063}

\bibitem[{{Skowron} et~al.(2014)}]{Skowron2014}
{Skowron}, D.~M., et~al., \emph{{OGLE-ING the Magellanic System: Stellar
  Populations in the Magellanic Bridge}}, \emph{\apj} \textbf{795}, 108 (2014),
  \eprint{1405.7364}

\bibitem[{{Skowron} et~al.(2016)}]{Skowron2016}
{Skowron}, D.~M., et~al., \emph{{OGLE-ing the Magellanic System: Photometric
  Metallicity from Fundamental Mode RR Lyrae Stars}}, \emph{\actaa}
  \textbf{66}, 269 (2016), \eprint{1608.00013}

\bibitem[{{Soszy{\'n}ski} et~al.(2015)}]{Soszynski2015}
{Soszy{\'n}ski}, I., et~al., \emph{{The OGLE Collection of Variable Stars.
  Classical Cepheids in the Magellanic System}}, \emph{\actaa} \textbf{65}, 297
  (2015), \eprint{1601.01318}

\bibitem[{{Soszy{\'n}ski} et~al.(2016)}]{Soszynski2016}
{Soszy{\'n}ski}, I., et~al., \emph{{The OGLE Collection of Variable Stars. Over
  45 000 RR Lyrae Stars in the Magellanic System}}, \emph{\actaa} \textbf{66},
  131 (2016), \eprint{1606.02727}

\bibitem[{{Udalski} et~al.(2015){Udalski}, {Szyma{\'n}ski}, \&
  {Szyma{\'n}ski}}]{Udalski2015}
{Udalski}, A., {Szyma{\'n}ski}, M.~K., {Szyma{\'n}ski}, G., \emph{{OGLE-IV:
  Fourth Phase of the Optical Gravitational Lensing Experiment}}, \emph{\actaa}
  \textbf{65}, 1 (2015), \eprint{1504.05966}

\end{thebibliography}

\end{document}